\documentclass[onecolumn]{article}
\begin{document}
\baselineskip 20 true pt

\def\w{{\omega}}
\def\dtauk{{\delta_k}}
\def\dtauj{{\delta_j}}
\def\Tjk{{\cal T}_k^{(j)}}
\def\Dw{{{\cal D}\omega}}
\def\bfw{{\bf w}}
\def\s{{\cal S}}
\def\slph{{\cal H}({\cal L};P_G^*,q)}
\def\r{{\bf r}}
\def\rr{{\cal R}}
\def\bfz{{\bf z}}

\title{Discrete Scale Invariance in the Cascade Heart
Rate Variability of Healthy Humans}

\author{D.C. Lin \\
Mechanical and Industrial Engineering Department\\
Ryerson University,Toronto, Ontario, M5B 2K3}

\date{\today}

\maketitle

\bigskip

\begin{abstract}
Evidence of discrete scale invariance (DSI) in daytime healthy
heart rate variability (HRV) is presented based on the
log-periodic power law scaling of the heart beat interval
increment. Our analysis suggests multiple DSI groups and a
dynamic cascading process. A cascade model is presented to
simulate such a property.
\end{abstract}

\bigskip

\section{Introduction}

The phenomenon of heart rate variability (HRV) in humans desrcibes
the beat-to-beat, apparently random, fluctuation of the heart
rate$^1$. HRV measured by the time span between ventricular
contractions, known as the beat-to-beat RR interval (RRi), is also
known to share many characteristics found in other natural
phenomena. For example, daytime RRi in healthy humans exhibits
1/f-like power spectrum$^2$, multifractal scaling$^{3,4}$, and
similar increment distribution observed in fluid turbulence$^4$.
These characteristics may vary significantly in heart disease
patients depending on the severity of the disease$^{1,5}$.

The origin and the generation of HRV remain the biggest challenges
in the contemporary HRV research. Although the respiratory and
vascular systems constantly modulate the heart rate, they do not
explain the large percentage of the broad-band (multifractal)
signal power in HRV. For example, it is unlikely that this broad-band
feature results directly from the output of the narrow-band
respiratory dynamics$^6$. Also, it is known that the level and the
variability of blood pressure and heart rate can change significantly
from upright to supine positions. In a 42-day long bed rest test,
Fortrat et al. showed that the variation in blood pressure and heart
before and after the test are qualitatively different, suggesting
separate control mechanisms for generating their variability$^7$. It
is thus believed that a more sophisticated structure may exist,
which integrates the feedback from receptors to create the pattern
of HRV$^8$.

Apart from its origin, some progess on the HRV generating mechanism
may be possible by using the discrete (lattice) multiplicative
cascade model$^4$. This is purely a phenomenology approach that
does not prescribe to any physiology term. Nontheless, encouraging
results were obtained that are consistent with the physiological
data in health and in certain heart disease$^4$. The main purpose
of this work is to investigate the basis of this modeling strategy.
Our approach is based on the scale invariant symmetry implied from
the HRV phenomenology$^{4,9,10}$. Since RRi cannot be defined
between heart beats, it is appropriate to consider discrete scale
invariance (DSI) in HRV. It is known that discrete cascade implies
DSI$^{11\sim 15}$. Better characterization of DSI in HRV is thus
important since it is the necessary condition for the multifractal
scaling observed in HRV. The existence of cascade is also
significant because it represents a very different approach of the
cardiovascular dynamical system from feedback control that is
additive in principle. The idea will support the previous studies
that a direct influence from baroreflex to multifractal HRV is
unlikely$^7$, as well as the need to search for a role by the higher
control centers in HRV$^8$.

The consequence of DSI is an oscillating scaling law with a well-defined
power law period. Such a scaling law is said to exhibit log-periodicity
(LP). In this work, we analyzed DSI in daytime healthy HRV by searching
LP in the scaling of HRV. Typically, LP is ``averaged out" in the process
of finding the scaling law. Using the technique called ``rephasing," this
problem can be effectively resolved and evidence of multiple DSI groups
in the healthy daytime RRi data was found. In light of this new result,
a cascade model is constructed using random branching law to reproduce
not only some ofthe known HRV phenomenology, but also the multiple DSI
characteristics.

The results of this work are organized in five sections. In Section 2, a
brief review of the notion of DSI is given. The numerical procedures for
identifying the DSI property from time series are described in Section 3.
Numerical examples and results on daytime heart rate data sets are given
in Section 4. Concluding remarks are given in the last Section.

\section{Discrete Scale Invariance and Rephasing}

\subsection{Ideas of Discrete Scale Invariance in Physical Systems}

A random processes $x(t)$ is said to possess continuous scale
invariant symmetry$^{13}$ if its distribution is preserved after
the change of variables, $t\to\lambda t$, $x\to x/\mu$ where
$\lambda$ and $\mu(\lambda)$ are real numbers; i.e.,
$$x(t)={1\over\mu}x(\lambda t).
\eqno(1)$$
DSI is defined when (1) only holds for a countable set of scale factors
$\lambda_1,\lambda_2,\cdots$. Scale invariance implies power law. The
power law in DSI has a log-periodic correction of frequency $1/\log(
\lambda)$: i.e., $x(t)=t^\gamma F(\log(t)/\log(\lambda))$ where $\gamma
=\log\mu/\log\lambda$ and $F(x)=F(x+1)$. Generally$^{15}$, one can
consider $x(t)=C_t t^{\gamma^\prime}$, $C_t$ being $t$-dependent, and
$\gamma^\prime=\gamma+2\pi n i/\log(\lambda)$ is a complex number for
$n=1,2,\cdots$. Novikov suggested LP in the small scale energy cascade
of the turbulent flow$^{14}$. Sornette and co-workers showed that LP
exists more generally in physical and financial systems, such as
turbulence$^{15}$, earthquake$^{16}$, rupture$^{17}$ and stock market
crashes$^{18}$.

The existence of the discrete scale factor implies a hierarchical
structure. This link can be simply illustrated by the middle
third Cantor set with the scale factor $\lambda=3$. With proper
rescaling, a precise copy of the set is only obtained with a
3-fold manification of the scale$^{13}$. If $x(t)$ denotes the
Lebesgue measure at scale $t$, the Cantor set can be modeled by
(1) using $\lambda=3$ and $\mu=\lambda^{-\log(2)/\log(3)}$.
Thus, the power law exponent of $x(t)$ (the box dimension of
the Cantor set$^{19}$) assumes a log-periodic oscillation of
frequency $1/\log(3)$ about its mean value $\log(2)/\log(3)$.

The hierarchical structure can be a dynamic object as a result of
some time-dependent branching law. Such a dynamic hierarchy is
believed to exist, for example, in the cascade paradigm of the energy
exchange in fluid turbulence where the break-down or ``branching"
of large-scale vortices into ones of smaller scales can occur
randomly in space-time with the energy re-distribution following
a multiplication scheme. In data analysis, the dynamic hierarchy
poses a technical difficulty for finding the scale factor $\lambda$
since LP may be averaged out in the process of obtaining the power
law. Zhou and Sornette proposed to conduct averaging {\it after}
rephasing or re-aligning data segments using a central maximum
criterion$^{15}$. Using this technique, these authors successfully
extracted LP in turbulence and proposed the DSI symmetry and
cascade. The rephasing technique is adopted in this work. Instead
of the central maximum criterion, the cross-correlation property of
the data segments will be used (see step (d) below).

\subsection{Rephasing RRi Data}

Let $\r(t)$ denote the RRi between the $t^{\rm th}$ and $(t+1)^{
\rm th}$ heart beats. Based on the turbulence analogy of HRV$^4$,
we focus on the LP in the scaling exponent of the empirical law
$\s(\tau,p)=\langle|\Delta\r(\tau)|^p\rangle\sim\tau^{\zeta(\tau,
p)}$ where $\Delta\r(\tau)=\r(t+\tau)-\r(t)$ and $p$ is a real
number. The implementation of the rephasing follows a 8-step
algorithm; see Fig.~1.

\bigskip
\noindent (a) Divide $\r(t)$ into $M$ nonoverlapping segments
$\{\rr_1,\cdots,\rr_M\}$.

\noindent (b) For $\r(t)\in\rr_i$, calculate $\s_i(\tau,p)=\langle
|\Delta\r(\tau)|^p\rangle$.

\noindent (c) Apply a low-pass $(K,L)$ Savitzky-Golay (SG)
filter$^8$ to $\s_i(\tau,p)$ and calculate its first derivative
to obtain a $\tau$-dependent $\zeta_i(\tau,p)$ for $i=1,\cdots,M$.
The $(K,L)$ SG filter performs a $K^{\rm th}$ order polynomial
fit over $L$ samples$^{15}$. It can produce a smoothing effect in
the high frequency while preserving the statistical moments of
the signal up to the order of the filter.

\noindent (d) Randomly select the $i$th segment as the base segment
and compute the cross-correlation $C_{i,j}^{K,L}(\kappa)$ between
$\zeta_i(\tau,p)$ and $\zeta_j(\tau+\kappa,p)$ for $j\ne i$.

\noindent (e) Shift the time origin of $\zeta_j(\tau,p)$ by $\Delta
_j$, where $\max(C_{i,j}^{K,L})=C_{i,j}^{K,L}(-\Delta_j)$, so that
the cross-correlation between $\zeta_i(\tau),p$ and the shifted
$\zeta_j(\Delta_j+\tau,p)$ has a maximum at zero time lag. Note that
$\Delta_i=0$ for the base segment.

\noindent (f) Average the shifted $\zeta_j(\tau+\Delta_j,p), j=1,
\cdots,M$, to obtain $Z_{K,L}(\tau,p)$.

\noindent (g) Compute the spectrum of $Z_{K,L}(\tau,p)$.

\noindent (h) Return to (c) with different $K,L$ values.

\bigskip
\noindent A Lomb periodogram$^{20}$ ${\cal L}(f)$ is used to estimate
the spectrum of $Z_{K,L}(\tau,p)$ for its superiority in handling
situations where noise plays a fundamental role in the signal, as well
as its capability in handling small data set.

Although the above algorithm provides the systematic steps to
estimate the log-periodic component, noise in the empirical data
can also generate spurious peaks in the Lomb periodogram. For
independent Gaussian noise process, this problem can be analyzed
by the false alarm probability$^{20,21}$:
$$P_G(f)=1-(1-\exp(-{\cal L}(f)))^m
\eqno(2)$$
where $m$ is proportional to the number of points in the spectrum.
The smaller the value $P_G(f)$ is, the more likely a genuine
log-periodic component exists in the signal. Thus, a Lomb peak with
large $P_G(f)$ suggests a chance event. Zhou and Sornette conducted
extensive simulations and showed that (2) is in fact an upper bound
for a number of correlated noise except for those showing long-term
{\it persistent} correlation$^{21}$. The fractional Brownian motion
(fBm) of a Hurst exponent greater than 0.5 is an example where (2)
does not apply. The multiple scaling exponents in healthy daytime
HRV have been found to lie below such a threshold$^{1,3,4,9}$ and we
will continue to use (2) in this work.

As shown above, DSI is characterized by the frequency $1/\log(
\lambda)$ of the LP. However, significant Lomb peaks may only
capture the higher harmonics $k/\log(\lambda)$, $k\ne 1$. It is
therefore necessary to define the relation of the significant
peaks. We propose a simple procedure to achieve this. First, we
collect the significant peaks satisfying $P_G\le P_G^*$ for $P_G
^*\ll 1$ and for different SG filter parameters. Second, we form
a significant Lomb peak histogram (SLPH) and locate its local
maxima. These maxima identifies the most probable frequencies of
the log-periodic oscillation of the power law. Let such maxima
be $f_1,\cdots,f_n$. The last step of the procedure is to search
the smallest $\lambda$ to minimize
$$d_\lambda=\sum_i^n |f_i-k_i/\log(\lambda)|
\eqno(3)$$
for integers $k_i$'s. We seek the smallest $\lambda$ since, with
finite precision in numerical computing, $d_\lambda$ can be made
arbitrarily small as $\lambda\gg 1$ This minimization step is
simple, easy to implement and, as we show below using synthetic
data, it is also effective.

\section{Numerical Results}

\subsection{DSI in Discrete Bounded Cascade}

The rephasing algorithm introduced above was first tested on 
synthetic data generated by the discrete cascade$^4$
$$x_J(t)=\Pi_j^J\omega_j(t)
\eqno(4)$$
where the cascade components $\omega_j(t)$ are discrete-time
processes given by
$$\omega_j(t)=1+\sigma_j{\bf w}
\eqno(5)$$
for $t_k^{(j)}\le t<t_{k+1}^{(j)}$, $k=1,2,\cdots$, $j=1,\cdots,
J$, and ${\bf w}$ is a zero-mean Gaussian random variable of
variance 1. Let $t_{k+1}^{(j)}-t_k^{(j)}=\delta_j$. The scale
factor $\lambda$ in the DSI hierarchy is related to $t_k^{(j)
}$'s by
$$\delta_j/\delta_{j+1}=\lambda.
\eqno(6)$$
To model the bounded RRi, we further assume $\sigma_j>\sigma_{
j+1}$ to assure boundedness. This model has been used in the
past to simulate HRV phenomenology, including transition of RRi
increment probability density function and multifractal
scaling$^4$.

According to (4), we generated 30 sets of dyadic ($\lambda=2$)
and triadic ($\lambda=3$) $x_J(t)$ with the corresponding
$\log(\sigma_j)=(-1.6-0.126j)\log(2)$ and $\log(\sigma_j)=(
-1.9-0.093j)\log(3)$, respectively. Each $x_J(t)$ has 8192
points and is divided into segments of 1024 points. Twenty-four
sets of $(K,L)$ SG filter are defined based on $K=3,4,\cdots,
7$, $L=7,9,\cdots,15$. For each combination of $K,L$, steps (c)
to (h) in the rephasing algorithm is repeated six times based
on six different base segments selected in step (d) of the
algorithm. This is implemented to avoid bias from a particular
segment. Significant Lomb peaks are collected based on the
false alarm probability $P_G^*<1\%$ or ${\cal L}(f)\ge 10$ and
$m=256$ points of the Lomb periodogram. The results for $p=2$
is reported as no quantitative difference exists for $p\le 3$.
Numerical results for $p>3$ show more variability due to poor
statistics.

FIG.~2a shows the $\zeta_i(\tau,p)$ of a particular segment of one
of the dyadic $x_J(t)$'s. The log-periodic oscillation with a
log-period $\log(2)$ is clearly seen. The Lomb periodogram of $Z_
{K,L}(\tau,p)$ (step (f) above) is shown in FIG.~2b based on a
particular choice of $K,L$ and the dominant LP is seen to pick up
the second harmonics of $1/\log(2)$. The SLPH estimated for
different SG filters over 30 sets of $x_J(t)$ is obtained in FIG.~3.
The clustering of the local maxima at integer multiples of $1/\log
(2)$ is evident. The minimization (3) identifies the correct scale
factor $\lambda=2$ for the dyadic cascade. Similar results of the
tradic cascade are also found (FIG.~3). These examples demonstrate
the effectiveness of the proposed numerical procedures.

\subsection{DSI in Daytime Healthy HRV}

For HRV, two databases are considered. The first set (DB1) consists
of 10 ambulatory RRi recordings from healthy young adults$^4$.
These test subjects were allowed to conduct normal daily activities.
The second set (DB2), available from the public domain$^{22}$,
consists of 18 ambulatory RRi recordings showing normal sinus
rhythm. The parameters used in the numerical analysis are the same
as above except the data segment length has increased to 2048
points. The choice of the segment length is a balance of two factors:
small segment length results in more segments but poorer statistics
in the estimation of $\zeta_i(\tau,p)$; large segment length results
in less segments but better estimate of $\zeta_i(\tau,p)$. We tried
1024 points per segment and found similar results; i.e., the group
averaged $\lambda$ value is similar to the ones reported in FIG.~5
below.

The SLPH in all cases shows well positioned local maxima that can
be easily related to the harmonics of some fundamental frequency
(FIG.~4). The $\lambda$ values for DB1 and DB2 are summarized in
FIG.~5. It is observed that (a) there are non-integer scale factor
$\lambda$ and (b) the $\lambda$'s cluster in the range of [3.5,
5.5] and the group averaged $\lambda$ are $\sim$4.8 and $\sim$4.4
for DB1 and DB2, respectively. The noninteger $\lambda$ unambiguously
excludes the possibility of discrete cascades with one scale
factor. It implies more complicated branching law and multiple DSI
groups in healthy HRV.

Although HRV and turbulence exhibit similar phenomenology$^4$, it
is interesting to point out the rather large $\lambda$ value ($>4$)
compared with the $\lambda\sim 2$ in fluid turbulence$^{10}$. From
the discrete cascade viewpoint, a larger $\lambda$ is compatible
with the ``patchiness" appearance commonly observed in the RRi of
healthy humans$^{1,3,4,9,10}$ since the $\omega_j(t)$'s of the
cascade will fluctuate on a longer time scale to create the effect.

To model the multiple DSI in cascade HRV, the scale factor $\lambda$
used in (5) is set to be a random number so that the log-periodic
oscillation of $\zeta(p)$ can vary over a range of frequencies. We
generated 30 sets of $x_J(t)$ according to (4) using uniformly
distributed $\lambda$ in the interval [2,6]. The simulated $x_J(t)$
exhibits the ``patchiness" pattern observed in the RRi data (FIG.~6),
and similar scaling characteristics found in the past$^4$ (FIGs.~7a
$\sim$ 7c). The scaling exponent $\zeta(\tau,p)$ of the power law
$\langle|\Delta x_J(\tau)|^p\rangle\sim\tau^{\zeta(\tau,p)}$
exhibits log-periodic oscillation that is captured by the well
positioned local maxima in SLPH (FIGs.~7d, 7e). In addition, the
average of the $\lambda$'s lies close to the group-averaged $\lambda$
values of DB1 and DB2 (FIG.~5).

\section{Conclusion}

It is known that discrete cascade leads to DSI and characterized
by log-periodic modulation of the scaling property$^{11,12}$.
Hence, the LP reported in this work supports the view of a cascade
for the multifractal generation in HRV. It implies a more
sophisticated process than reflex-based control mechanisms that
function on the additive basis. It also suggests the need to
search for a role by the higher control centers in HRV$^8$. It is
conjectured that the cascade describes the process which
integrates the regulatory feedbacks in the cardiovascular system
to create the pattern of HRV.

The non-integer scale factor implies multiple DSI. This property
was also reported in the screening competition of the growth of
diffusion limited aggregation model$^{23,24}$. To the best of our
knowledge, this is the first instance of multiple DSI being
reported in HRV. We do not have the better knowledge of its origin,
except to believe it reflects the multiple time-scale control
mechanisms in the cardiovascular dynamical system.

It is tempting to search for the physiological correlate of the
cascade, for example, the role of the cascade components $\omega_
j(t)$. Based on the spectral analysis, we suggested that the
large time scale components ($\omega_j, j\sim 1$) capture mainly
the sympatho-vagal interaction and the small time scale components
($\omega_j,j\gg 1$) capture the parasympathetic activity$^4$.
However, we should caution that cascade is a modeling tool
derived from statistical physics. The $\omega_j(t)$ can therefore
represent the range of micro- to macroscopic processes in the
cardiovascular dynamical system.

A rather narrow range of the scale factor $\lambda\in[3.5,5.5]$
estimated from the two different databases implies a ``stable"
hierarchical structure of the cascade that does not vary sensitively
with the details of the healthy population. The analysis of the
identified DSI characteristics in other physiological conditions is
currently underway and its result will be reported in the near
future.

\bigskip
\noindent{\bf Acknowledgment}
\bigskip

This research is supported by Natural Science and Engineering Research
Council of Canada. The author would like to thank many years of
valuable comments and suggestions by Dr. R.L. Hughson of the University
of Waterloo and critical comments by the anonymous referee.

\vfill\eject
\noindent{\bf Reference}
\bigskip

\noindent [1] Task Force of the ESC and NASPE, {\it Euro. Heart J.}, 
{\bf 17}, 354 (1996).

\noindent [2] M. Kobayashi and T. Musha, {\it IEEE Trans. Biomed. Eng.},
{\bf 29}, 456 (1982).

\noindent [3] P.CH. Ivanov, et al., {\it Lett. to Nature}, {\bf 399},
461 (1999).

\noindent [4] D.C. Lin and R.L. Hughson, {\it Phys. Rev. Lett.}, {\bf 86},
1650 (2001); D.C. Lin and R.L. Hughson, {\it IEEE Trans. Biomed. Engng.}, 
{\bf 49}, 97 (2002); D.C. Lin, {\it Fractals}, {\bf 11}, 63 (2003); D.C.
Lin, {\it Phys. Rev. E}, {\bf 67}, 031914 (2003).

\noindent [5] G.C. Butler et al., {\it Clin. Sci.}, {\bf 92}, 543 (1997).

\noindent [6] Y. Yamamoto et al., {\it Am J. Physiol}, {\bf 269}, H480 (1995);
G.C. Butler et al., {\it Am J Physiol}, {\bf 26}, R26 (1994).

\noindent [7] J.O. Fortrat et al., {\it Auton. Neurosci.}, {\bf 86}, 192 (2001).

\noindent [8] F. Togo and Y. Yamamoto, {\it Am. J. Physiol. Heart
Circ. Physiol.}, {\bf 28}, H17 (2001).

\noindent [9] H.E. Stanley, et al., {\it Physica A}, {\bf 281}, 60
(2000); P.Ch. Ivanov, et al., {\it Chaos}, {\bf 11}, 641 (2001).

\noindent [10] P. Bernaola-Galvan, et al., {\it Phys. Rev. Lett}, {\bf 87},
168105 (2001).

\noindent [11] F. Schmitt and D. Marsan, {\it Eur. Phys. J. B}, {\bf 20}, 3-6 (2001).

\noindent [12] E. Bacry, J. Delour, and J.F. Muzy, {\it Phys. Rev. E}, {\bf 64}, 026103 (2001).

\noindent [13] D. Sornette, Critical Phenomenon in Natural Sciences,
Series in Synergetics, {\it Springer Verlag}, Heidelberg (2000).

\noindent [14] E.A. Novikov, {\it Dokl. Akad. Nauk SSSR}, {\bf 168}, 1279 (1966).

\noindent [15] W-X. Zhou and D. Sornette, {\it Physica D}, {\bf 165}, 94 (2002);
D. Sornette, {\it Phys Rep}, {\bf 297}, 239 (1998).

\noindent [16] A. Johansen, et al., {\it J. Geophys. Res.}, {\bf 105},
28111 (2000); Y. Huang, et al., {\it J. Geophys. Res.}, {\bf 105},
28111 (2000). 

\noindent [17] Y. Huang, et al. {\it Phys. Rev. E}, {\bf 55}, 6433 (1997).

\noindent [18] A. Johansen and D. Sornette, O. Ledoit, {\it J. Risk}, {\bf 1},
5 (1999).

\noindent [19] K. Falconer, Fractal Geometry, Mathematical Foundation and
Applications, {\it John Wiley \& Sons}, Chichester (1990).

\noindent [20] J.D. Scargle, {\it Astrophys. J.}, {\bf 263}, 835 (1982).

\noindent [21] W-X. Zhou and D. Sornette, {\it Int. J. Mod. Phys. C}, {\bf 13}, 137 (2002).

\noindent [22] A.L. Goldberger, et al., {\it Circulation}, {\bf 101}, e215 (2000).

\noindent [23] A. Arneodo, et al., {\it Phys. Rev. Lett}, {\bf 68}, 3456 (1992).

\noindent [24] D. Sornette, et al., {\it Phys. Rev. Lett}, {\bf 76}, 251 (1996).


\vfill\eject
\noindent {\bf Figure Captions}

\bigskip
\noindent FIG. 1 Sketch of the numerical procedure for rephasing.
The second segment is illustrated as the base segment and
rephasing was shown for $\zeta_1(\tau+\Delta_1,p),\cdots,\zeta_M
(\tau+\Delta_M,p)$ ($\Delta_j$ is determined at the maximum of
the cross-correlation function between the $j$th and the base
segments). Log-periodicity in $Z_{K,L}(\tau,p)$ is estimated from
the Lomb periodogram.

\bigskip
\noindent FIG. 2 (a) $\zeta_i(\tau,p) - \langle\zeta_i(\tau,p)
\rangle$ versus $\log(\tau)$ taken from the synthetic dyadic
bounded cascade. The solid line is a pure sine wave with a
period of $\log(2)\sim 0.693$. (b) Typical Lomb periodogram of
$Z_{K,L}(\tau,2)$ (averaged over all $\zeta_i(\tau,p)$'s).

\bigskip
\noindent FIG. 3 SLPH estimated from 30 sets of (a) synthetic
dyadic bounded cascade $x_J(t)$ and (b) triadic $x_J(t)$. The
grid lines in (a) and (b) are drawn according to $k/\log(2)$
and $k/\log(3)$, $k=1,2,\cdots$, respectively.

\bigskip
\noindent FIG. 4 (a) SLPH of a typical data set from DB1. The
local maxima $f_{\max}$ are marked by ``+". (b) $f_{\max}$
versus $k/\log(4.5)$, $k=1,2,\cdots$, showing as the harmonics
generated by the fundamental frequency $1/\log(4.5)$. The
straight line has the slope $1/\log(4.5)$. (c) Similar to (a)
based on a data set taken from DB2. (d) Similar to (b) based
on the local maxima of (c). The straight line has the slope
$1/\log(3.1)$. Note the local maximum between $7/\log(3.1)$
and $8/\log(3.1)$ was not fitted by the harmonics of $1/
\log(3.1)$.

\bigskip
\noindent FIG. 5 Scale factor $\lambda$'s for 10 subjects in
DB1, 18 subjects in DB2 and 30 sets of synthetic data $x_J
(t)$ generated by the cascade model. The group averaged
$\lambda$ values and standard deviations are superimposed
and drawn as ``$\bullet$" and vertical bar, respectively.

\bigskip
\noindent FIG. 6 A typical sample of $x_J(t)$ (top) and the
RRi data (bottom) taken from DB2. Both time series show the
characteristic of ``patchiness" in their fluctuation pattern.

\bigskip
\noindent FIG. 7 (a) to (c) show the $1/f$-like power spectrum,
power law, $\s(\tau,p)$ and the nonlinear $\zeta(p)$ of $\s(
\tau,p)$, respectively, of the $x_J(t)$ shown in FIG.~5; see
Ref. 4 for the similar characteristics reported for RRi data in
healthy humans. (d) and (e) show the SLPH of two typical $x_J(
t)$. Well-positioned local maxima in (d) and (e) capture the
harmonics generated by $\lambda$: $\sim$4.4 and $\sim$3.85,
respectively.

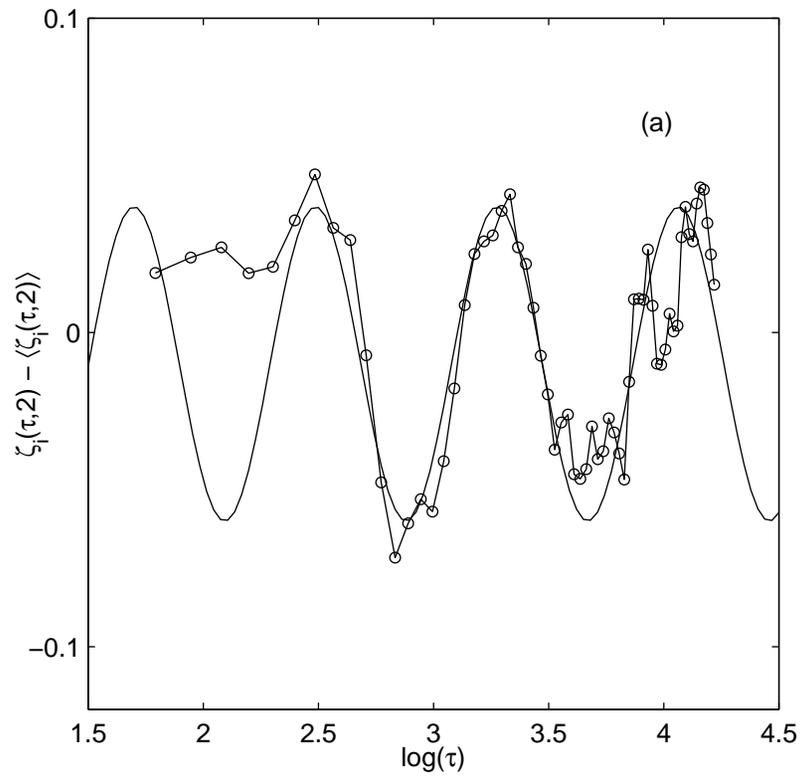
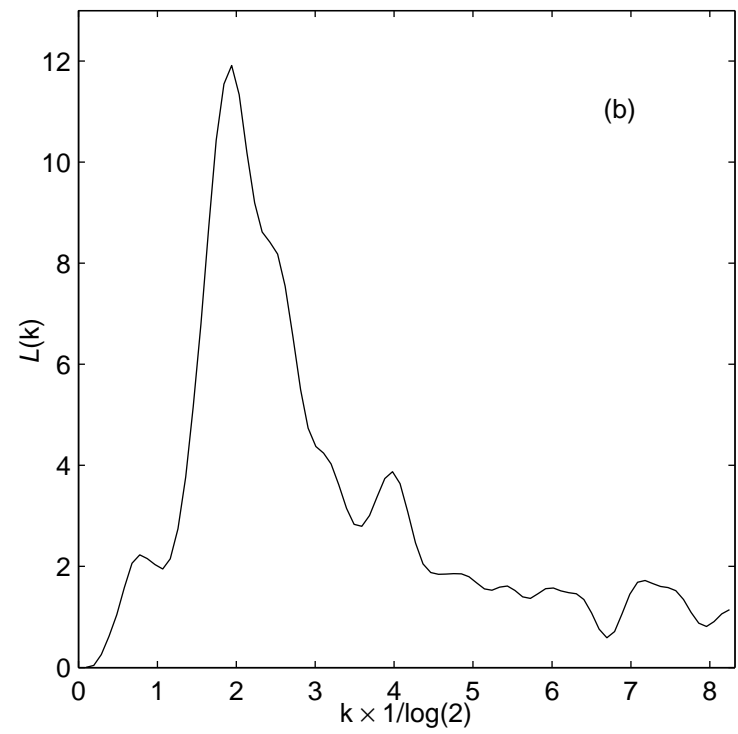
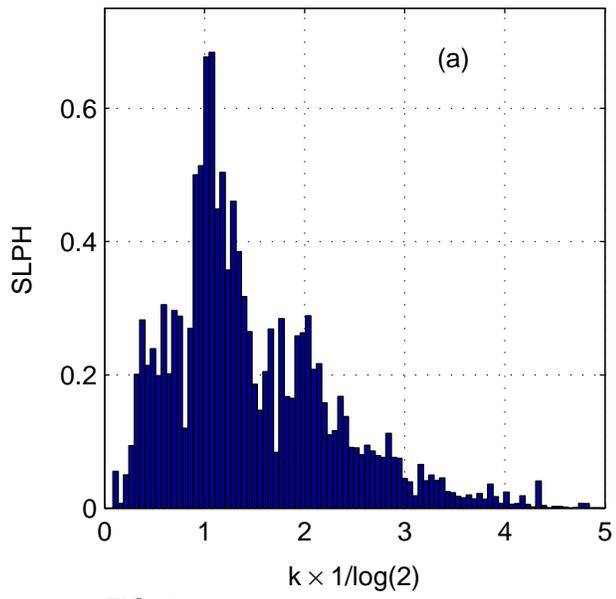
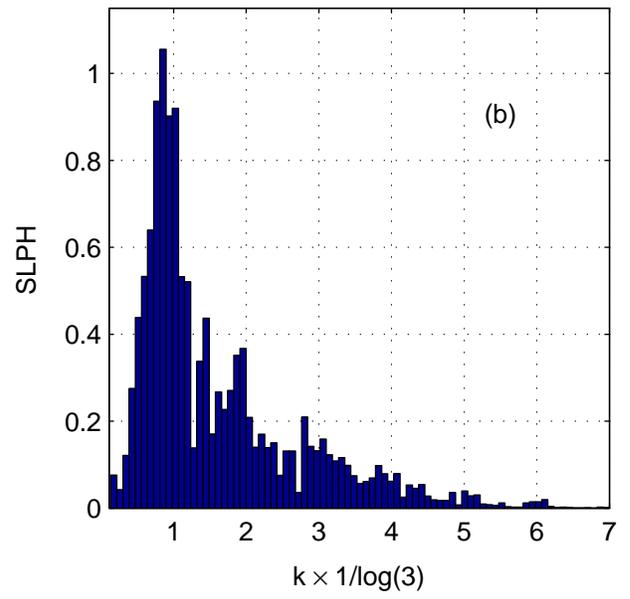
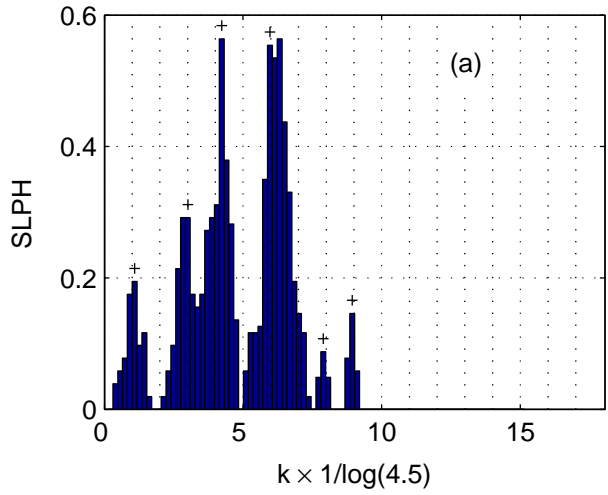
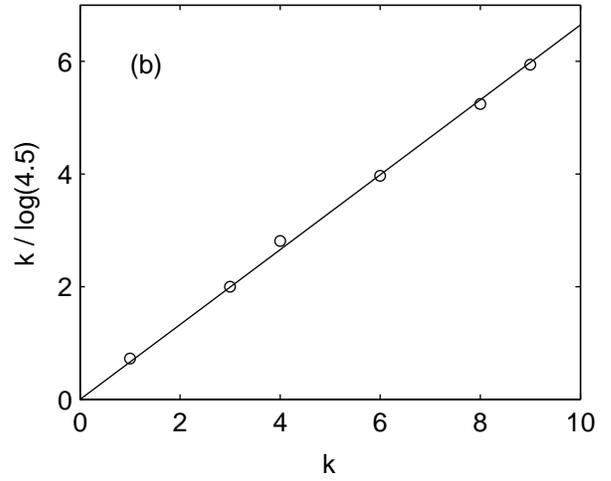
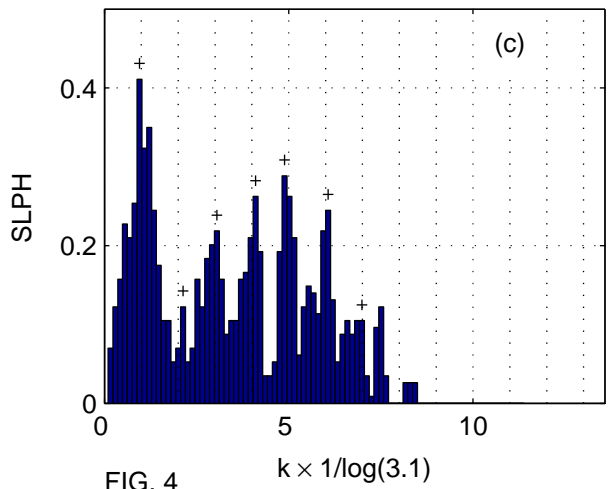
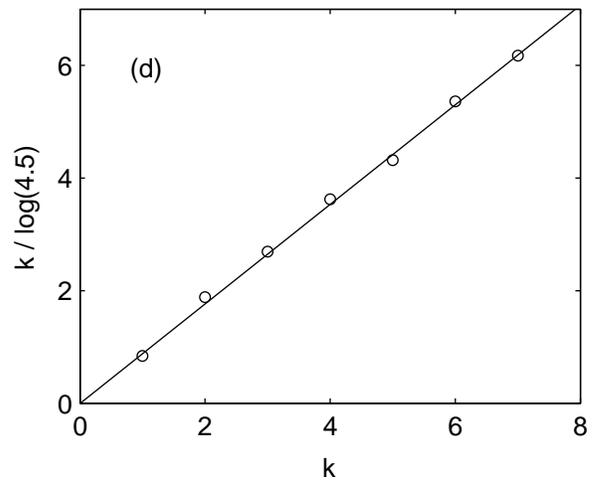
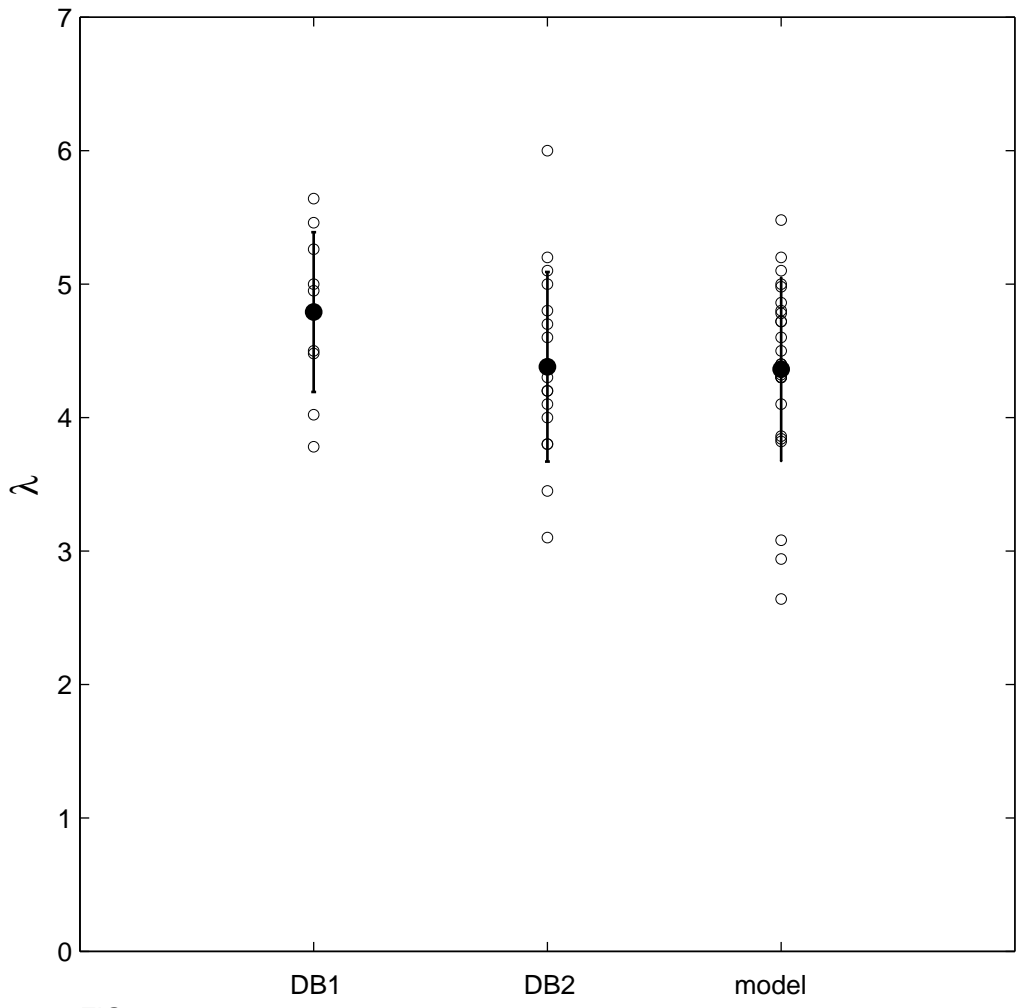
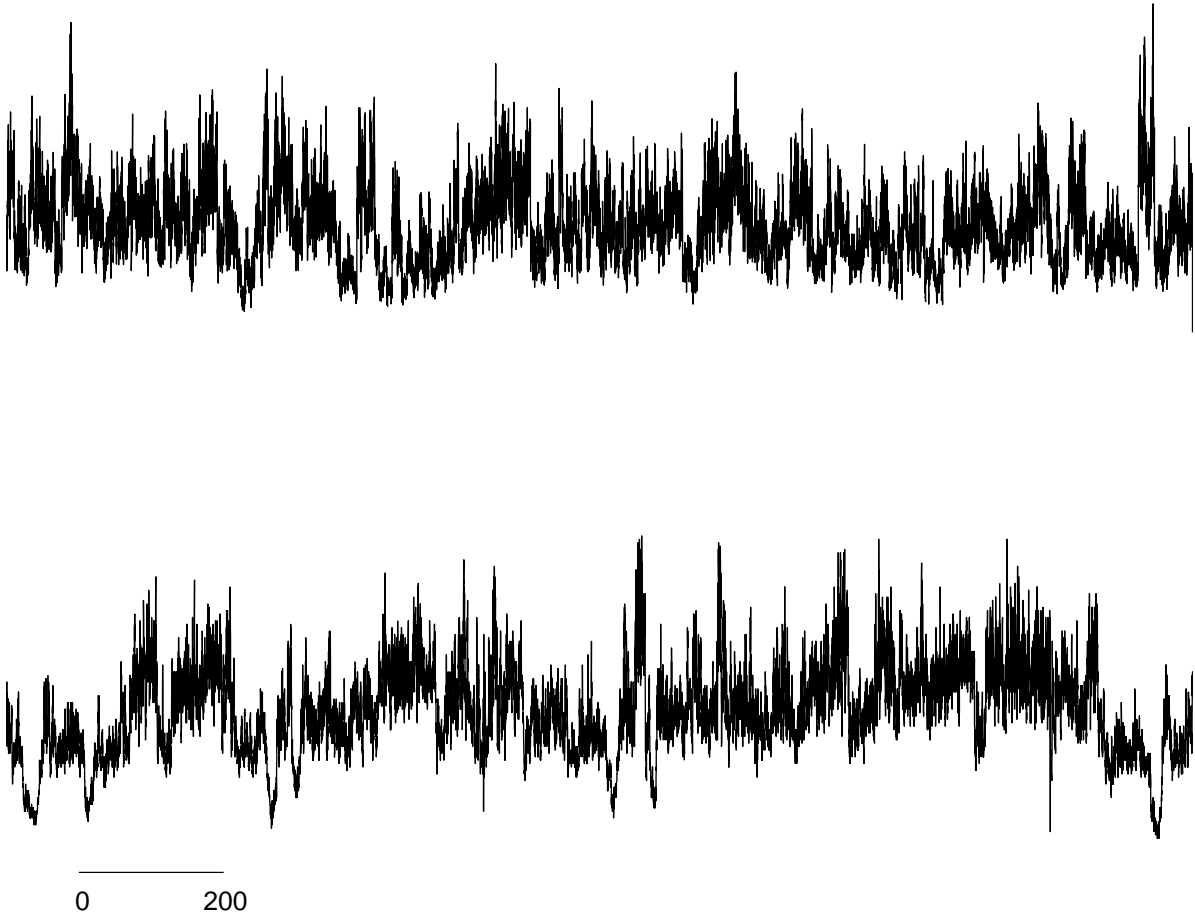
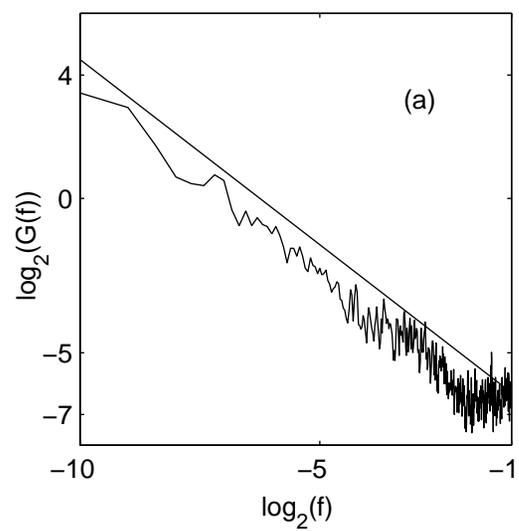
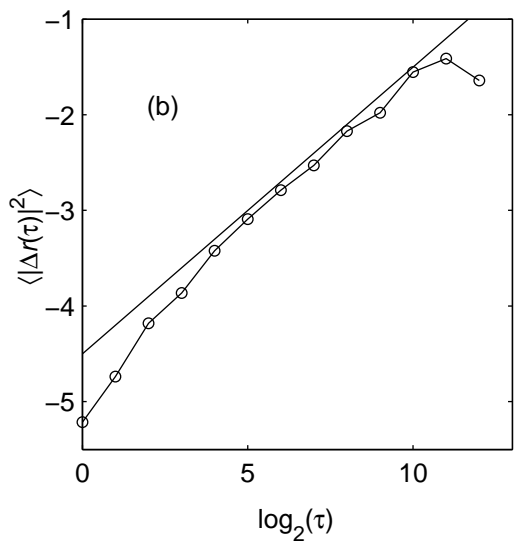
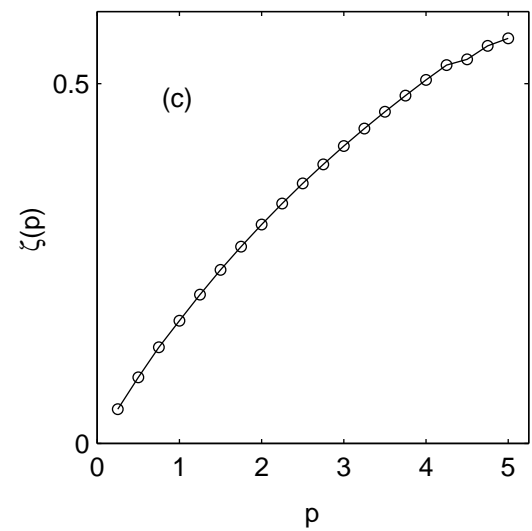
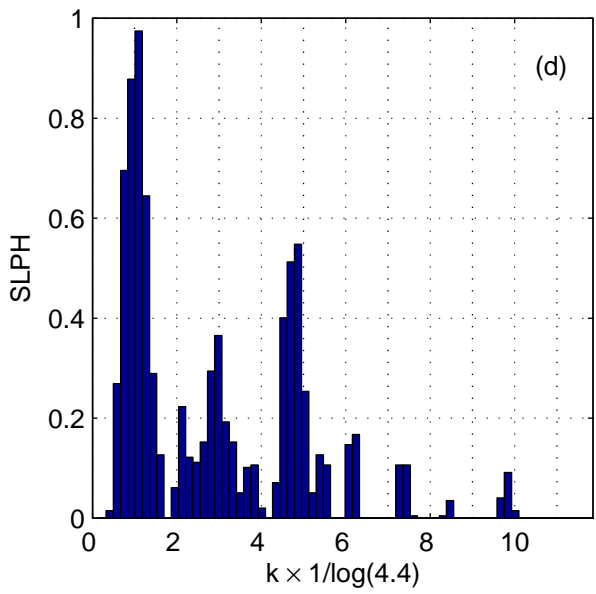
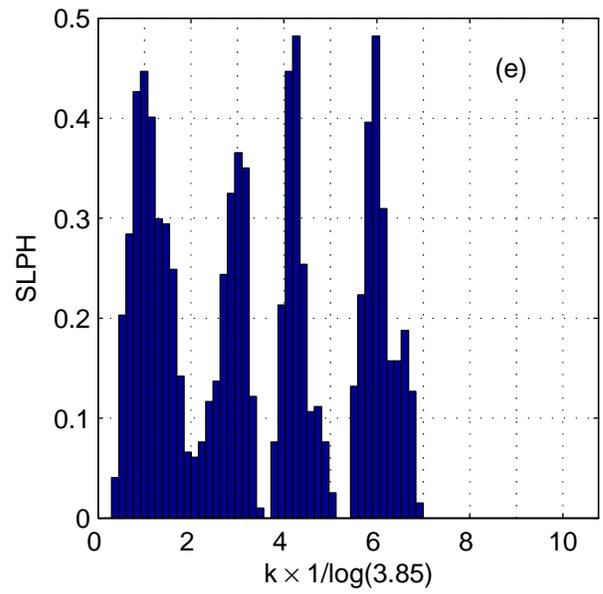
\end{document}